\documentstyle[a4,12pt]{article}
\def\bea{\begin{eqnarray}}  \def\eea{\end{eqnarray}}
\def\1{{\rm 1\mskip-4.5mu l} }
\begin{document}

\begin{center}
\vspace*{1 truecm}
{\Large \bf Reply to Aharonov and Anandan's ``Meaning of the} \\
{\Large \bf Density Matrix"} \\[8mm]  
{\large \bf B. d'Espagnat}\par
{Laboratoire de Physique Th\'eorique et Hautes
Energies\footnote{Laboratoire
associ\'e au
Centre National de la Recherche Scientifique - URA D00063
}}\\ {Universit\'e
de Paris-Sud, B\^atiment 210, 91405 Orsay Cedex, France\\ Fax : 33 1 69 15 82 87 -
e-mail : espagnat@qcd.th.u-psud.fr} \end{center}

\vskip 1 truecm
\begin{abstract}
Aharonov and Anandan's claim that the notion of ``proper mixture'' is improper is
shown to be unjustified. The point is made that if a purely operationalist standpoint
is taken the three difficulties these authors describe relatively to the conventional
interpretation of density matrices in fact vanish. It is noted that nevertheless it
is very difficult for us to do without any form of realism, in particular when the
quantum measurement problem is considered, and it is stressed that the proper mixture
notion comes in precisely at this level. The more general question of the real
bearing of Aharonov and Anandan's ideas on the interpretation of quantum mechanics
problem is considered. \end{abstract}

\vskip 2 truecm
\noindent LPTHE Orsay 98-28 \par
\noindent April 1998 \par

\newpage
\pagestyle{plain}

\section{Introduction} \hspace*{\parindent}

Aharonov and Anandan recently issued an article entitled {\it Meaning of the Density
Matrix} \cite{1r} in the main text of which the ideas of a certain physicist called
d'Espagnet are criticized. The first interesting question this paper raises is, of
course: ``Who is this man~?''. Admittedly the idea occurred to me that, conceivably,
he could be I. But there were tokens of the contrary. The recurrent spelling mistake
was admittedly but a minor one. A more serious indication was the following.
According to the text, this d'Espagnet had tried to ``point to the state of a physical
system that is represented by $\rho$ in the sense that measurements on this state
would give $\rho$''. Since the authors wrote this, I thought, it must be true. And
since personally I never made any such attempt (I do not even understand the authors'
description of it...)  Mr. d'Espagnet could not possibly be I. Paradoxically, this
(optimistic) impression of mine was enhanced by the fact that, in the text,
d'Espagnet is attributed by name a view actually shared, I believe, by, practically,
all quantum physicists, namely the idea that, $|\phi_{\alpha}>$ and $N_{\alpha}$
being, respectively, the normalized state vector describing an ensemble $E_{\alpha}$
of physical systems and the number of elements in the latter (all systems being of the
same type, and $N = \Sigma_{\alpha}N_{\alpha}$) the operator

\begin{equation}
\rho = \Sigma_{\alpha}\Big | \phi_{\alpha} \Big |{N_{\alpha} \over N}
\Big |<\phi_{\alpha} \Big | \label{1e}
\end{equation}

\noindent describes a mixture (incidentally, in the authors' formula (4), here
relabelled (\ref{1e}), the symbol $\Sigma_{\alpha}$ is missing). The authors refer to
some drawbacks they claim this view has and, in fact, I have come to suspect that, at
least in some parts of the article, this d'Espagnet is just a convenient symbolic
figure representing the large set of the physicists who naively entertain the said
view without even noticing its drawbacks. However I well realize that this cannot be
the whole story. After all, the name ``d'Espagnat'' is correctly spelt in both the
abstract and the references and... I {\it did} write the book entitled {\it Veiled
Reality} \cite{2r}! So I finally decided that I had to consider the substance of the
problem raised. It is the subject matter of this note.
	
One of my claims concerning it is that, in fact, this problem has much to do - much
more than is apparent at first sight - with the difficulty of setting a consistent
relationship between two notions, that of {\it operational definition} (of systems and
their properties) and that of ontological existence. Hence, in Section 2 the point is
made that if a purely operationalist standpoint is taken the three difficulties
Aharonov and Anandan' Introduction describes in fact vanish. In Section 3 it is noted
that nevertheless it is very difficult for us to do without any form of realism, in
particular when the quantum measurement problem is considered, but that difficulties
then creep in; and it is shown that, in this field at least, contrary to Aharonov and
Anandan's claim, the notion of proper mixtures is justified. In Section 4 the more
general question of the bearing of Aharonov and Anandan's ideas is investigated and
in Section 5 some conclusions are drawn.

\section{On the virtues of unaltered operationalism} \hspace*{\parindent}

The necessity of only using concepts having a well defined meaning prompts us to
systematically use operational definitions but we have a natural trend towards
realism and in quantum physics reconciling the two is not always easy. One possible
standpoint is to resist this ``ontological'' trend and take but the operational
aspects of quantum physics seriously. It is then considered that quantum mechanics is
merely a set of rules correctly predicting what the outcomes of measurements will be.
Since some of these rules only yield probabilities we have to consider ensembles. But
it must be observed that, to repeat, the standpoint in question removes at one stroke
the three drawbacks to the use of ensembles that Aharonov and Anandan mention in
their Introduction. This is obvious concerning drawback (1). It is also clear
concerning drawback (2) since within this standpoint $\rho$ is but a predictive tool.
That a predictive tool should change abruptly when new information is gathered is
quite natural and does not constitute a conceptual problem. And drawback (3) vanishes
as well since within the said standpoint introducing the ensemble in question - a
Gibbsian one, that is, an abstract concept - only serves to give a meaning to the
probabilities of getting such and such measurement results. The formalism simply
{\it does not} include the rule that the result obtained in one measurement should be
relevant for predicting the probabilities of further measurement outcomes. Rather, it
stipulates that to this end a new Gibbsian ensemble corresponding to $\rho$ should be
imagined afresh.

\section{Charms and Dangers of Realism. Proper\break \noindent mixtures}
\hspace*{\parindent}

The above conclusion is clear and, I think, uncontroversial. Wherefrom does it then
come that Aharonov and Anandan found the conventional meaning given to density
matrices has the drawbacks they state? This question admits of but one answer. It is
because just as all ``normal'' people, they do not strictly cling to ``pure
operationalism and nothing more''. Implicitly or not, they instill in their views some
admixture of realism.

This, of course, is fully rational and understandable. To appreciate how natural it
is the best way is to consider a problem different from theirs, namely the quantum
measurement problem. Suppose we have to do with an ensemble of generalized,
Schr\"odinger-cat-like measurements, made with instruments having scales with $\mu$
intervals and taking place before time $t$. If we want to strictly remain within the
realm of a purely operationalistic description we must be very cautious concerning
the way we describe the state of affairs after time $t$. We must do this just by
stating that if we look at the pointers after that time we shall get the ``feeling''
that some of them lie in scale interval 1, some of them lie in scale interval 2 etc.,
the corresponding proportions being derivable from knowledge of the pointer ensemble
density matrix (obtained by partial tracing from the density matrix of the overall
ensemble of systems plus instruments). In a ``strictly scien\-ti\-fic'' sense this
information incorporates everything that we need to know. A more ``realistic''
description of the pointers, supposing that it can be given, would add nothing to our
predictive powers concerning the phenomenon of interest. But still, most people find
it very difficult to believe that this operational description is the ``whole
story''. Quite naturally they would like to be able to interpret the ensemble $E$ of
all the pointers ``realistically'', that is, as composed of $\mu$ subensembles
$E_1,\cdots ,E_{\alpha},\cdots ,E_{\mu}$, the components of each $E_{\alpha}$  being
pointers really lying in one definite interval, the one labelled $\alpha$. This shows
that, at least in some situations, it is hardly possible to resist our natural
inclination towards the philosophical standpoint called realism. What differentiates
this realistic description from the foregoing, purely operational one is that we now
consider that on every member of each $E_{\alpha}$ the result of measuring in which
interval it lies is predetermined. This assumption cannot be translated in a purely
operationalistic language since it is impossible to operationally ascertain it. To be
sure we can measure in what interval this pointer is but nothing proves that it is
not this very measurement act that sets it in this interval. Nevertheless the realist
considers the assumption in question as meaningful an he finds it so natural that he
deems it to be correct.

But then, as is well known, a problem arises. As soon as, following most of the
theorists who tried to build up quantum measurement theories, we decide to describe
quantum mechanically - that is, either by state-vectors or by density matrices - each
one of these $E_{\alpha}$ ``as it really is'' (i.e. taking into account the fact that
each one of its components is in scale interval $\alpha$ and nowhere else), we
automatically get that since $E$ must be the addition of all the $E_{\alpha}$, it is
of the nature of what, by definition, I called a ``proper mixture'', describable by a
$\rho$ of the form (\ref{1e}) or a trivial generalization thereof. This simply follows
from the fact that, obviouly, under the conditions stated the quantum description of
any one of these $E_{\alpha}$ must differ from those of the others. Note that this
conclusion is of a general nature: the observation that macroscopic systems such as
pointers are not isolated from the environment does not alter it since we can, by
thought, incorporate environment within the system.  For completeness sake let it
here be briefly recalled that the conclusion in question may be viewed as being at
the source of the ``measurement problem'', that, within conventional realism, the
problem in question has no satisfactory solution, and that this is considered by some
as compelling us to give up the said realism. This analysis, however, fall outside our
present subject. \par \vskip 5 truemm

Concerning the present debate two points, I think, emerge from the above. One is that
physicists should more thoroughly investigate what {\it they} intuitively mean when
they, explicitly or implicitly, deal with realistic notions. Let this question be
deferred to Section 4. The other point is that, at least as it is used in my book
{\it Veiled Reality} \cite{2r}, the notion of ``proper mixture'' is not in the least
``improper'', contrary to Aharonov and Anandan's claim. The reason is that, as
explained in Remark 5, Section 7.3 of {\it Veiled Reality}, the chapters of this book
that Aharonov and Anandan refer to were written just for the purpose of making clear
the difficulty realism leads to. The whole argumentation in these chapters is, in this
respect, negative. In substance, it amounts to exclaim: ``Look here: in general
ensembles produced this way {\it are not} proper mixtures. You cannot think of them as
you would think of balls distributed among different vessels''. Considered in this
light, that is, as essentially linked to an {\it interpretation} problem - and more
precisely to attempts at building up an interpretation couched in realist, or
``classical'', terms - the notion of proper mixture far from being ``improper'' is
trivial. It is no more improper that the just introduced notion of balls distributed
among different vessels.

Let us turn, then, to Aharonov and Anandan's objections. First of all, it is true, of
course, that picking up one ball in one definite vessel changes the probabilities
concerning further pickings that might be done. But then what~? Does this make the
notion of ``balls distributed among vessels'' inconsistent, illogical, ``improper''?
Obviously not. The same holds true here. What must be stressed in this respect is
that the notion of balls distributed among vessels (or pointers distributed among
scale intervals) is one that we, originally, {\it have}; that, as we saw, we
intuitively tend to use it, at least in some instances, for giving a realistic
meaning to partial trace density matrices; and that, therefore, it was necessary to
show that it is not, in general, appropriate for this purpose. Otherwise said,
Aharonov and Anandan's remark about the ``memory'' of proper mixtures is correct but
irrelevant. It does not constitute an objection to the concept of such mixtures
since, as shown above, the concept in question is necessary for discussing
interpretations of quantum measurement theories that seem natural, not to say
``obvious''.

Hence their objection must boil down to one of a semantical nature; one of the type:
``the concept d'Espagnet calls `proper mixture' is a valid one but the ensemble it
refers to are not mixtures''. Since names always are, to some extent, conventional
such semantical discussions are not, as a rule, of much interest. But anyhow, my
opinion on this is that the arguments in favor of calling such ensemble ``mixtures''
are at least as cogent as those for abstaining of doing so. One of them is, of
course, that the word ``mixture'' is an element of our commonsense, everyday language
and that, conceptually, proper mixtures are considerably more similar to such
``everyday life'' mixtures than are the mixtures I call ``improper'' (that is, the
ones for which Aharonov and Anandan insist on saving up the word `mixture'). Another
and even more significant argument is that, for all purely operational purposes such
as those described here in Section 2, proper mixtures can be described by density
matrices, that is, are indistinguishable from the ensembles for which Aharonov and
Anandan save up the word presently under discussion (indeed, it is only when `proper'
mixtures are given some kind of an ontological interpretation, as collections of
physical systems, that differences with `improper' mixtures appear, see below and
Ref. \cite{2r}).

As for the other criticisms that Aharonov and Anandan try to develop in their Section
4, I think they are unsubstantial as well. In the paragraph in which these authors
let $N$ tend to infinity they observe that the density matrix (\ref{1e}) cannot be any
one of ``this sequence''. Unfortunately, in their wording it is not clear what the
expression ``this sequence'' actually means. They also observe, as previously noted,
that so long as only usual measurements are performed on the system ``we cannot point
to the state of a physical system that is represented by $\rho$ in the sense that
measurements on this state would give $\rho$''. But in what sense is this a
criticism~? It seems that Aharonov and Anandan have here in mind their own view that
a density matrix should be attached to a single system, not realizing that it is not
necessarily the view other physicists have in mind.

By contrast, their analysis of the differences I had noted between differently
prepared proper mixtures described by the same density matrix is basically sound. In
subs\-tance, however, it is a mere rewording of what I had already stated in Section
8.3 of Ref.~\cite{2r}. The points I made there were {\it (i)} that the argument
``these two mixtures are different since they have been produced differently'' is not
a universally convincing one; {\it (ii)} that while observable differences (the ones
Aharonov and Anandan mention) can indeed be produced between these mixtures, the
differences in question are actually observable - hence significant - only if these
``ensembles'' are treated as what, after all, they physically are, that is, as
systems of (noninteracting) particles; and finally {\it (iii)} that concerning the
question whether or not we can define a state by means of a non-pure-case density
matrix the answer is {\it yes} but only at a (heavy) price. This price, as I explained
there, consists in deciding that the ensemble concerning which quantum mechanics
yields predictions (concerning nonprotective measurement results, of course!) are
essentially abstract cons\-tructs, very useful indeed for predicting observation
outcomes but not to be considered as composite physical systems (i.e. not to be
considered as one would like to consider ensembles of, e.g., pointers, see above).
And, correlatively, that - at least in quantum measurement theory - the notion
``state'' is but a predictive concept.

At first sight this last point seems to contradict Aharonov and Anandan's views since
these authors consider with favor the concept of an ``objective reality that could be
described by the density matrix''. It is clear, however, that in their approach such
an objective reality is defined only relatively to protective measurements, and it
remains to be seen whether and to what extent such a partial definition of reality is
truly satisfactory. This is the purpose of the next section.

As a last remark concerning proper mixtures let me stress once more that, at the
start, two very general standpoints are conceivable, namely (i) the purely
operational, purely predictive one and (ii) the descriptive, tentatively realistic
one. A priori either one of these two standpoints is worth consideration but we have
to choose between them. If we choose the first one, or if our investigations lead us
to consider that it is the only a posteriori tenable one, then it is clear from the
above that there is no point in considering proper mixtures. But if we choose the
second one - or before we have finally discovered this second one is untenable - I
cannot really see why a {\it general} notion of proper mixtures (not restricted to
objects we would like to think of as classical) should be an inconsistent one. The
point is that within this second standpoint we cannot cling to the rule - call it
Rule R - that by definition two systems or ensemble of systems that cannot be
distinguished from one another by any measurement are, by definition, identical. This
rule R is specific to standpoint (i). If, in the spirit of standpoint (ii), we impart
by thought some reality to the quantum states there is then no reason why we should
not consider ensembles of systems of the same type lying in different quantum states.
As we saw above on the particular example of pointers, there are cogent reasons to
describe such ensembles by density matrices such as (\ref{1e}) and call them mixtures.
And it is easily seen that those of the objections raised by Aharonov and Anandan that
are valid can be disproved, just by observing that, implicitly, they are based on
Rule R and that, in standpoint (ii), Rule R is not present.

\section{Invariants and Reality} \hspace*{\parindent}

As we saw in Section 2, it is a fact that the purely operational standpoint works. On
the other hand, as pointed out in Section 3, it is also a fact that most people are
expecting more from physics. Indeed, many of us consider that the real purpose of
physics is to describe in detail the physical world {\it as it really is}.  And the
very wording of quantum physics - with such terms as ``particle'' and ``state'' -
illustrates how powerful this ideal is. Needless to say, it is a respectable one. It
motivates most of the inquiries concerning interpretation of quantum physics and in
particular the tentative quantum measurement theories. Concerning wave functions and
density matrices it explicitly constitutes the substance of Aharonov and Anandan's
aim. In my view the ideal in question (call it, say, the ``realist'' one) is not to be
dismissed on the basis of purely philosophical arguments. Indeed, it is one that, in
the high times of classical physics, was considered by most scien\-tists as being very
much within reach and it is reasonable that contemporary physicists with such a
realist turn of mind should do their best to preserve it. However, in Section 3 we
also touched upon several points that have been quite extensively developed by many
authors (see e.g.\cite{2r}) and show that, when the phenomenon of quantum measurements
is duly taken into account, the ideal in question generates considerable difficulties.
Hence the question really arises: the quantum formalism being what it is, is it
actually possible to salvage, within it, at least the main elements of a ``normal
kind'' of realism? This, to repeat, is the aim not only of the many authors of quantum
measurement theories but also, as it seems, that of Aharonov and Anandan as well.

To study this question it is appropriate to proceed more or less as philosophers
would. When discussing such matters, philosophers use to begin by introducing the
notion of {\it invariants}, a substantive referring to invariance with respect to
varied modes of apprehension and/or description. In ordinary life we have - they note
- visual etc. impressions that change when we move around: but our mind is able to
build up the notion of a set of {\it things} and {\it properties of things} (such as
shape, color etc.) that, we posit, do {\it not} really change under these conditions;
that are ``invariants'' with respect to them. Correlatively - they go on noting - our
mind spontaneously builds up a theoretical model making it possible to consistently
explain its impressions by referring to the invariants in question. And, they claim,
it is such invariants that, in ordinary life, we call real. Note in this respect
that, in this context, ``to call'' is more than just to introduce a semantical
convention. To call these invariants {\it real} means hypostatizing them to a kind of
special level. It means instinctively attributing to them an ``absoluteness'' that, as
we already noted, cannot be defined operationally (I mean: noncounterfactually) but
that somehow makes them differ from a host of other concepts, taste, equations,
probablities, etc. that are not viewed as ``physically real''.

Hypostatizing concepts goes together with hypostatizing the models that are based on
them. We then say, or would like to say, that the models in question are true
descriptions of the World as it really is. At this stage, however, an important point
must be stressed. It is that this whole hypostatizing procedure is really
satisfactory only if the validity of the hypostatized concepts and model is not
limited to the description of phenomena taking place within some limited experimental
context.  For further reference let us call this condition, {\it Condition A}. When
classical physics was in its apex Condition A seemed to be met. The World then
appeared as composed of particles and fields and the nature of each one of these
constituents could be specified quite independently of the particular phenomenon or
experimental procedure the physicists choosed to discuss. In other words, the
corresponding invariants were totally noncontextualistic. It was not necessary to
strictly associate any one of them with some restricted class of impressions. In that
sense, they could be termed universal.

If we set aside the so called ``ontologically interpretable models'' such as the
Broglie-Bohm model - which raise difficult problems of their own, would call for a
separate discussion (see e.g.\cite{2r}) and lie anyhow outside the realm of the
present debate - we must ackowledge that the advent of quantum theory dramatically
altered the above picture. Within Bohr's approach, for example, complementarity means
that the validity of any one of the ordinary language invariants is, in the
microscopic domain, limited to the description of phenomena taking place within some
well-defined experimental context, that is, to a restricted class of observables. In
other words, these invariants do not meet Condition A.

Did more recent advances in the field substantially modify the situation? I do not
think so. Take, say, the concepts of {\it events} and {\it histories}. It is true that
the theories of Griffiths, Gell-Mann and Hartle and Omn\`es partly succeeded in
restoring the meaningfulness of these notions within the microscopic domain. It is
true that, for example, within a given consistent history branch an event is, in
these theories, independent in principle of whether it is observed or not that is, it
qualifies for being considered as an invariant in this respect. However, it is not an
invariant with respect to the adjunction - or, better to say, the ``taking into
account'' - of some possible future events. In other words, it is an invariant only
with respect to quite a limited class of possible experimental procedures: a class
that is very far from incorporating all the experimental procedures that are easily
available. Hence it is impossible to hypostatize it to the level of an element of
reality as classical realists did concerning the invariants they thought were
universal. It does not fulfill Condition A.

Now, since this article is motivated by the Aharonov and Anandan paper, it is
worthwhile to point out that, same as in the foregoing example, the Aharonov and
Anandan concepts of wave functions and density matrices attached to one system only
are, as stressed for example by Bitbol \cite{3r,4r}, merely {\it partial} invariants.
They are invariant with respect to protective measurements but only with respect to
them.  Hence the above conclusion also holds good concerning them. These invariants do
not fulfill Condition A. Admittedly we may, if we like, take up the convention of
calling them real. Viewed as a mere convention this one may be useful (see below). On
the other hand, since protective measurements constitute a highly limited class of
possible experimental procedures, when trying to predict ordinary measurement
outcomes we shall obviously continue being in need of considering probabilities,
statistical ensembles and so on. In this domain, density matrices attached to one
system only will be of no help. Willy-nilly we shall have to go on interpreting them
as referring to the ensembles in question.  

 \section{Conclusion} \hspace*{\parindent}

There is no doubt that the protective measurements concept and the possibility it
yields of - in some conceptual contexts - interpreting density matrices as
descriptions of one system ``states'' are quite interesting indeed. This is shown for
example by the content of Section 3 of Aharonov and Anandan's paper. Of course, we
have known for a long time that as long as we decided to forget about such things as
the measurement problem and the Born probability rule, the remaining quantum rules
could be stated as genuine physical laws, that is, without referring to human actions
and observations: otherwise said, in a strongly objective language parallelling the
strongly objective language of classical physics. With this proviso we could say:
``there exist wave functions and/or density matrices that have such and such
properties, obey such and such equations etc.'', much as had been done in classical
physics concerning material objects and fields. But there remained the difference
that in classical physics the entities thus said to exist (positions of objects,
field strengths etc.) could also be measured, a circumstance that contributed very
much to remove any suspicion that their postulated existence was ``unwarranted
metaphysics''. This is a difference that the new ideas brought in by Aharonov and
Anandan remove. And in this sense it must be granted that these ideas do bring
quantum mechanics somewhat closer to a realistic world view that many physicists
consider as being the only acceptable one.

But on the other hand we should keep in mind that all this holds good only as long as
the proviso is accepted of forgetting about the measurement problem while, of course,
we cannot {\it really} forget about it. It is a fact that the great majority of the
measurement we perform are not ``protective'' ones, that they yield definite values
according to definite probability laws etc.. and it seems clear that, as stressed at
the end of the foregoing section, the existence of these effects will oblige us to
continue using probabilities, ensembles and density matrices in the usual way. Now,
in this field all the analyses that have been made, along the years, of the quantum
mechanical rules and the measurement problem (including the fact that improper
ensembles are not to be identified with proper ones!) remain valid. All taken
together, they show that the strongly objective language of the main parts of
classical physics cannot be consistently used in quantum physics. Finally therefore,
as shown in \cite{2r}, quantum physics as a whole can only be expressed in a weakly
objective language, in which objectivity is identified with universal
intersubjectivity. 

 \end{document}